\documentclass[epj]{svjour} 

\usepackage{epsfig,graphicx,amsmath} 
\usepackage{psfig,natbib} 
\usepackage{float,pst-all,cuted} 
 
\renewcommand{\=}{~=~} 
\newcommand{\be}{\begin{equation}} 
\newcommand{\ee}{\end{equation}} 
\renewcommand{\=}{~=~} 

\begin{document}

\title{Electromagnetic response and break-up of light weakly-bound nuclei  
in a dicluster model} 

\author{L.Fortunato and A.Vitturi} 
\institute{Dipartimento di Fisica "G.Galilei", Universit\`a di Padova and  
INFN,\\ via Marzolo 8, I-35131 Padova, Italy} 
 
\titlerunning{Break-up of dicluster nuclei}


\abstract{
The light weakly-bound nucleus $^7$Li is studied within a dicluster $\alpha+t$ 
picture. Different observables obtained within our simple model are  
compared  with previous calculations and experiments showing  
good agreement. In particular we calculate dipole and quadrupole  
electromagnetic response to the continuum.  
The energy distribution of B(E$\lambda$) values 
are consistent with the energy weighted molecular sum rule and display 
a sizable contribution of non-resonant character arising from the  
weak binding property. 
The corresponding form factors for excitations to the continuum are 
used in a 
semiclassical coupled channel scheme to get estimates for the break-up cross 
section in a heavy ion reaction. The nuclear contribution is found to play an 
important role in the process for bombarding energies around the Coulomb 
barrier. 
The masses and charges ratios of the two clusters are shown to lead to 
features of the cluster halo that may significantly differ from 
the one usually associated with one-nucleon haloes.
}
\maketitle

\section{Introduction} 
A distinctive feature of nuclear systems along the neutron drip-line 
is the concentration of multipole strength at excitation energies  
just above the continuum threshold.  
This concentration of strength (mainly of dipole or quadrupole nature)  
is directly  
measured in breakup reactions, but it has strong dynamical effects also  
on other processes, such as elastic scattering or sub-barrier fusion 
reactions. 
It has been proved that this peculiar feature is  
associated with the weakly bound nature of most nuclei at the drip-line  
(\cite{Cata,Das2,Hama,Alka,Typel}). 
Within a di-cluster description of a weakly-bound nucleus (where 
in the simplest case one of the clusters may be a single nucleon) 
the quantum state 
that describes the system lies very close to the threshold for separation into 
the two subsystems. The wavefunction for the relative motion 
associated with such a state (and 
hence its distribution of matter) extends to large radii, spreading far  
outside the walls of the intercluster potential well (this is valid  
already at the level 
of a square well potential, and it is even more evident for a realistic 
potential with a diffused surface). 
This establishes the opportunity to set a matching between the bound wave  
function and some scattering state in the (low-lying) continuum  
whose typical wavelength roughly corresponds to the spatial extension 
of the bound state wavefunction. As a consequence  
the resulting electromagnetic response 
shows a marked concentration of strength in the threshold region at  
an excitation energy directly correlated to the binding energy.
With the specific scaling that depends on 
the angular momentum of the initial state, 
as well as on the neutron or proton character of the halo state, 
the energy corresponding to the maximum of the strength  
distribution follows approximately a linear behaviour 
on the binding energy \cite{Naga}, 
while the total dipole strength at the threshold depends approximatively 
on the inverse of the binding energy and tends therefore to magnify its  
effects as one approaches the drip-lines. 

The picture outlined above finds its simplest application in the case of 
single particle haloes \cite{Das1,Cata}, 
where, in a mean field approach, it is the last 
unpaired nucleon that is responsible for the halo distribution, but it
can be extended to  
the case of light weakly-bound dicluster nuclei to describe excitations  
to continuum states that lead to cluster breakup.
A number of experiments have been pursued in recent years, for instance, 
on the study of break-up of Li isotopes \cite{Kelly,exp1,exp2,exp3}.
We will take as a paradigmatic example the case of the nucleus $^7$Li,  
whose ground state is well described in terms 
of interacting $\alpha$ and triton clusters, which characterize the lowest 
continuum threshold (at 2.467 MeV). The basic necessary  
assumption is that the excited states, both bound and unbound, are also 
described within the same dicluster picture, assuming the two clusters 
to be frozen. The excitation process 
is therefore reduced to a transition in the wave function describing the 
cluster-cluster relative motion. 

This simple model for the threshold strength is modified when the system  
displays, in the low-energy continuum, true resonant states in addition to  
the non-resonant part. This is for example precisely the case of 
$^7$Li which has the 
$7/2^-$ and $5/2^-$ states at 4.652 MeV and 6.604 MeV respectively.  
Within the cluster picture these states 
correspond to narrow resonances in the relative motion with angular momentum 
$\ell =3$. {\it Ad hoc} formalisms, which only include either the resonances 
or the non-resonant continuum, may therefore be inadequate to describe 
the full process. In a proper treatment of the response to the 
continuum both resonant 
and non-resonant contributions arise in a natural way and may have comparable 
strengths. As an example of such an approach we recall the recent  
work of Kelly and collaborators \cite{Kelly}, who 
analyzed experimental data within a CDCC approach \cite{cdcc}, in which the  
continuum is discretized. The binning of the continuum is however not
optimal due to computational limitations, a problem that as we will see, 
is not present in our model. CDCC calculations usually consider only a few
energy bins in the relevant low-energy region, while our approach can
easily accomodate for thousands of bins in the same energy range.
Other CDCC calculations for $^7$Li breakup
are found in Ref.\cite{Keel}. 

In our calculation,  
the form factors for excitations to the continuum have been used in a 
semiclassical coupled channel scheme to get estimates for the break-up cross 
section.  As an example we have chosen the specific reaction 
$^7$Li$+~^{165}$Ho 
for which subbarrier fusion data are available and for which estimates of 
break-up probabilities are important for the interpretation of the data  
\cite{indian}. Since there are indications that the nuclear field play
a non-negligible role \cite{mason},
both Coulomb and nuclear contributions are included and their relative  
importance is analyzed. The non-resonant contribution to the cross section is 
found to provide 
a sizable fraction of the total cross section. Due to the strong 
nuclear component, the quadrupole break-up process is predicted to dominate  
over the dipole. 
Since optical parameters for holmium are not available, we have also tested 
our model against other calculations \cite{Kelly} for the reaction 
$~^7$Li + $~^{208}$Pb, finding good agreement.

\section{Dicluster description of $^7$Li} 
    
Walliser and Fliessbach \cite{Wall} discuss a  
cluster picture for $~^7$Li, in which the constituents of the nucleus 
(the $\alpha$ and $t$ particles) 
are treated as elementary, that is without internal structure, but not  
necessarily point-like. They obtain considerable agreement with 
experimental data and we conform, in principle, to their model.  
The main difference is the choice of the potential to be used to 
determine the relative motion of the cluster. 
In similarity with the usual single-particle case, our  
effective $\alpha-t$ potential  
\be 
V_{\alpha-t}(r)\=V_{coul}(r)+V_{WS}(r)+V_{\bf l \cdot s}(r) 
\ee 
contains, besides the coulomb repulsion (corrected at small distances for the  
sphericity of charge distributions), the nuclear attractive  
potential (assumed of simple Woods-Saxon form) and the  
spin-orbit term \cite{Bohr}. 
The depth of the Woods-Saxon well ($V_{WS}=-74.923$ MeV) and  
the magnitude of the spin-orbit correction ($V_{ls}=1.934$ MeV) have been  
adjusted 
to reproduce the energy eigenvalues for the two bound states. 
The $\alpha$ cluster has spin $0$ while the $t$ cluster has spin ${1\over 2}$. 
The angular momentum coupling between the $\ell =1$ relative motion  
and the spin of the 
triton provides the total angular momenta $({3\over 2})^-$ for the ground  
states with energy $-2.467$ MeV and $({1\over 2})^-$ for the first excited  
state at $-1.989$ MeV \cite{Till} (Energies are measured with respect  
to the $\alpha-t$ break-up threshold).  The resulting wave functions 
for the ground state and 
for the first excited state are in a qualitative 
agreement with the ones obtained in the paper of Wallisser and Fliessbach  
(for example the radial node occurs at the same point). 
 
\begin{table}[t] 
\begin{center} 
\begin{tabular}{|l|p{1cm}|c|p{1.4cm}|} 
\hline 
Quantity&This work&Experiments&Other works\\ \hline \hline 
$<r^2>_{ch}^{1/2}(fm)$&
$2.44$&
$2.55 (0.07)^{13}$&
$2.43^{10,18}$\\ 
&&$2.39(0.03)^{13}$&$2.55^{19}$\\ \hline 
$Q_{el} (fm^2)$&$-3.77$&$-3.8 (1.1)^{13}$&\\ 
&&$-3.4 (0.6)^{13}$&\\ 
&&$-3.70 (0.08)^{13}$&\\ \hline 
$Q_{mat} (fm^2)$&$-3.99$&$-4.1 (0.6)^{13}$&
$-3.82^{10}$\\ 
&&$-4.00 (0.06)^{16}$&$-3.83^{10}$\\  
&&&$-4.41^{10}$\\ \hline 
$B(E2,{3\over 2}^-\rightarrow{1\over 2}^-)$&$7.55$&$8.3 
(0.6)^{13}$&$7.74^{10}$\\ 
$(e^2fm^4)$&&$8.3 (0.5)^{13}$&$7.75^{10}$\\ 
&&$7.59 (0.12)^{16}$&$10.57^{18}$\\ 
&&$7.27 (0.12)^{16}$&\\ \hline 
$B(M1,{3\over 2}^-\rightarrow{1\over 2}^-)$&$2.45$&$2.50 (
0.12)^{13}$&\\ 
$(\mu^2)$&&&\\\hline 
$\Gamma ({7\over2}^-)(keV)$&$\sim 110$&$93 (8)^{17}$&\\ 
$\Gamma ({5\over2}^-)(keV)$&$\sim 930$&$875^{+200}_{-100}$&\\ \hline 
\end{tabular}\\ 
\end{center} 
\caption{Comparison of calculated and experimental quantities. The second  
column shows our results, while the third are various experimental data. 
The last contains calculations performed by other authors. The apices in 
parenthesis indicate the references.} 
\end{table} 
 
In spite of its simplicity, this model for $~^7$Li is nevertheless capable of 
a good agreement with experimental observations, as witnessed by the list of 
observables in Table 1. Evaluation of charge radius, electric and matter  
quadrupole moments, $B(E2)$ and $B(M1)$ values for transitions between  
the ground state and the first excited state are reported. 
These quantities, except the two width, are calculated accordingly  
to the prescriptions given in Ref. \cite{Wall}. 
These observables are very sensitive to the particular shape of the 
wavefunctions and therefore 
provide a reliability test for our approach as far as bound states are  
concerned. In the same table we  also compare our findings with previous 
calculations. The last two rows  
in the table refer to the widths of the two $f_{7/2}$ and $f_{5/2}$  
resonances which are given with the purpose to show that this model  
gives also sensible predictions for the continuum states.

\section{Electromagnetic response} 
 
We now apply the dicluster picture to the calculation of  
electromagnetic response for the transitions 
to continuum states. In this scheme all the features of the  
transition are ascribed to the modification of the wavefunctions 
describing the relative motion. The clusters are in fact assumed to be 
 frozen in this picture, and their intrinsic wavefunctions are not  
modified by the electromagnetic operators. 
\begin{figure}[!t] 
\begin{center} 
\epsfig{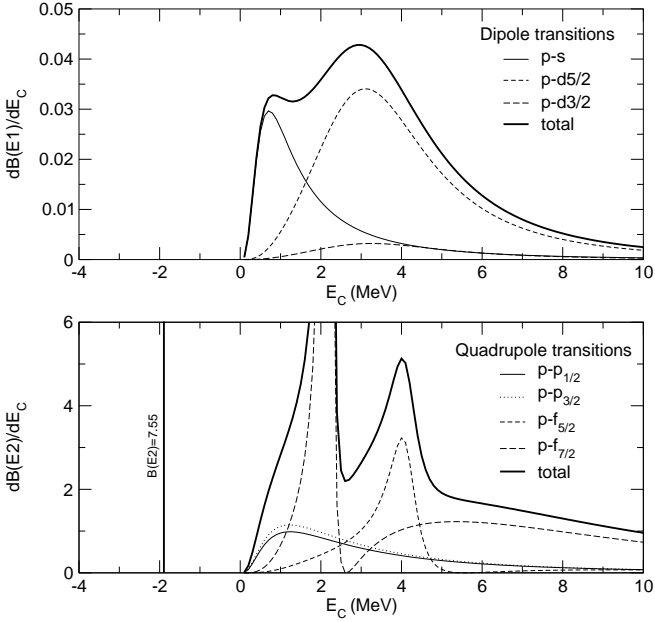}  
\end{center} 
\caption{Upper panel: differential B(E1) values (in $e^2 fm^2/MeV$) 
for transitions from the ground state to the continuum.  
Energies are in MeV, the different contributions are indicated 
in the legend. Lower panel: differential B(E2) values (in $e^2 fm^4/MeV$) 
for transitions from the ground state to the continuum and to the first 
excited bound state, displayed in the figure as a discrete bar (the transition
strength is indicated). Energies are in MeV, referred to the  
threshold for break-up into the $\alpha-t$ channel. } 
\label{dipqua} 
\end{figure}  
The multipole operator may be written as a sum of operators that 
act on different degrees of freedom. Since there is no rearrangement of 
the intrinsic structure of the two clusters the corresponding parts will 
not contribute to the $B(E\lambda)$.  
The strength distribution for the transition from an initial state of the 
intrinsic motion with wavefunction $\psi_{n_il_i}(r)$ and quantum numbers 
$n_i,l_i,j_i$ to a different final state (either bound or unbound) with  
wavefunction $\psi_{n_fl_f}(r)$ and quantum numbers $n_f,l_f,j_f$ 
may be written as 
$$ 
B(E\lambda) \= {\hat{j_f}^2 \hat{l_f}^2 \hat{l_i}^2  
\hat{\lambda}^2\over 4\pi}~ e^2_\lambda 
  \Biggl({l_f\atop 0}{\lambda \atop 0} {l_i \atop 0}\Biggr)^2  
\Biggl\{{l_f\atop j_i}{j_f \atop l_i} {j_{cl} \atop \lambda}\Biggr\}^2 $$ 
\be 
\cdot \Biggl( \int_0^\infty \psi_{n_fl_f}(r) r^{\lambda+2} 
\psi_{n_il_i}(r) dr \Biggr)^2 \,,
\ee 
where $\hat{j}=(2j+1)^{1/2}$ and the effective charge is defined as  
$e_\lambda \= Z_{cl}(A_{co}/ A)^\lambda+ 
Z_{co} ( -A_{cl} / A)^\lambda$  
and the subscripts $cl$ and $co$ refer to the cluster (the one with a nonzero  
intrinsic angular momentum) and the core (the one with a null intrinsic  
angular momentum). When the final state is in the continuum, its wavefunction  
also depends on $E_C$.

Starting from the ground state (with $p$ character) we have investigated  
electric dipole transitions to $s$ and $d$ states   
as well as quadrupole transitions to $p$ and $f$ states. The corresponding 
differential transition probabilities are shown in figs. (\ref{dipqua}).  
In the former case the scattering states for even  
multipolarities have been calculated   
with the same potential that has been used to generate the bound states. 
In the latter case the same parameters have been used for the  $p$-wave  
continuum, while for the $f-$wave modified Woods-Saxon  
($V_{WS}=-68.255$) and  spin-orbit ($V_{ls}=3.115$) potentials  
have been used to yield the $7/2^-$ and $5/2^-$ resonant states in the  
excitation spectrum at the correct energy. With this choice 
the widths of  
these two states have been found in reasonable agreement with  
experimental observations as shown in the last part of table 1, without  
the need for further adjustments. 
Besides this resonant strength we observed a concentration of strength of 
non-resonant character at the separation threshold, solely due to the  
weakly-bound nature of the $~^7$Li nucleus. This strength is small for  
multipolarities that have a resonance in the low-lying continuum, but it is 
sizable when there are no resonances  (as in the $p$ cases). 
\begin{figure}[!t] 
\begin{center} 
\begin{picture}(300,160)(0,0) 
\psset{unit=1.pt}   
\scalebox{.8}{
\pscircle*(280,100){20}   \rput(280,60){Target} 
\pscircle(80,50){35}\rput(80,10){Core} 
\pscircle(100,140){20}\rput(100,165){Cluster} 
\psline{<->}(80,50)(100,140) 
\psline{->}(280,100)(80,50) 
\psline{->}(280,100)(100,140) 
\psline{->}(280,100)(91,100) \rput(80,100){CM} 
\rput(130,108){$\vec R$} 
\psbezier[linewidth=0.5]{->}(40,180)(100,130)(30,140)(94,112) 
\rput(35,190){$ f_1 \vec r$} 
\psbezier[linewidth=0.5]{->}(10,60)(40,70)(20,98)(89,88) 
\rput(00,50){$ -f_2 \vec r$} 
\rput(180,140){$ \vec R+f_1 \vec r$} 
\rput(180,60){$ \vec R-f_2 \vec r$} }
\end{picture} 
\end{center} 
\caption{Coordinate system for the interaction between a dicluster nucleus 
(white) and an external target (black).} 
\label{coord} 
\end{figure}
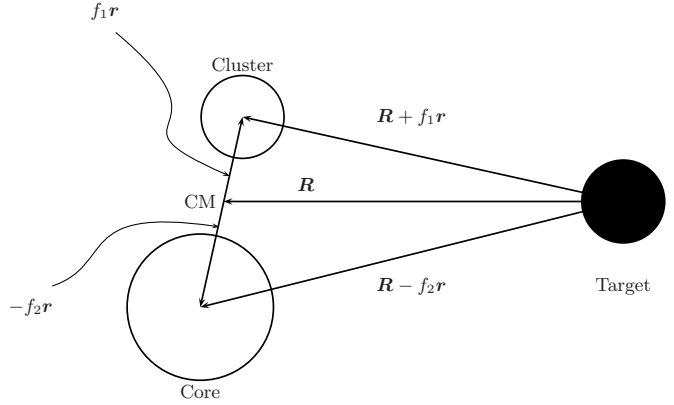     
 
\begin{figure*}[!t]
\epsfig{figure=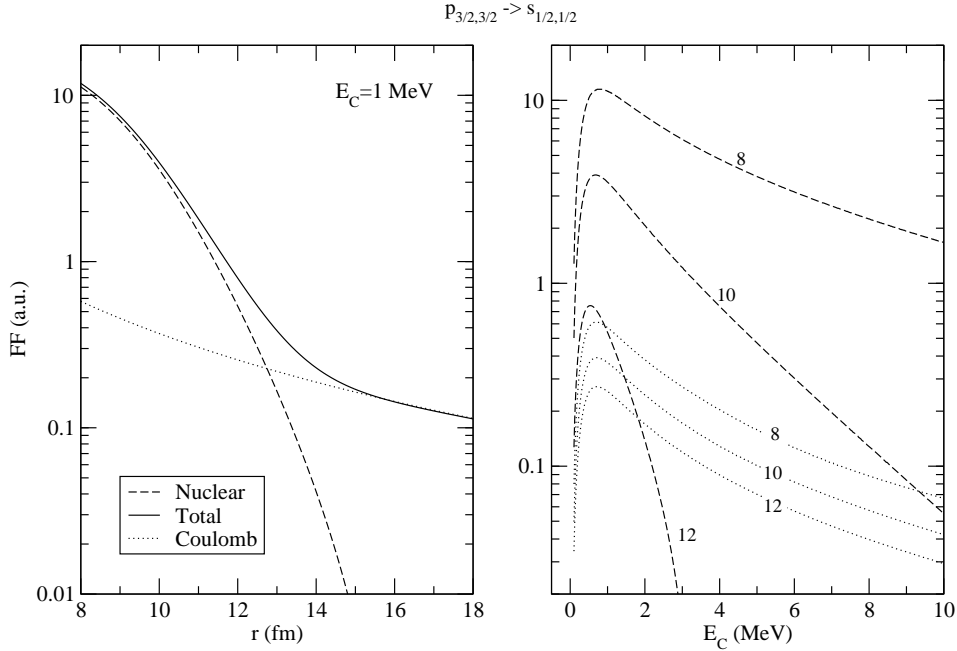,width=.7\textwidth} 
\caption{Form factors (in arbitrary units) for a particular transition  
plotted against the distance, for a fixed energy in the continuum of  
$E_{C}=1$ MeV (left panel, logarithmic vertical scale)  
and against the energy in the continuum for 
three fixed distances (right panel). Coulomb (dotted) and nuclear (dashed) 
form factors are shown. See text for details.  } 
\label{formfact} 
\end{figure*} 
We have compared our calculated values with the predictions of the  
energy weighted sum rules as well as 
of the energy weighted molecular sum rules (EWMSR) \cite{Alha,Lang}, also  
called AGB sum rule, that are particularly useful for  
molecular-like structures. In light nuclei enhanced $E1$  
transitions have been observed for which $B(E1)$ values may still be very 
small 
in comparison with single-particle estimates. EWMSR have been introduced 
as a measure for these transitions and in the cases of dipole and quadrupole  
they read: 
\be 
S_I(E1,A_1+A_2)= \left( {9\over 4\pi} \right) 
{(Z_1A_2-Z_2A_1)^2 \over AA_1A_2} \left( {\hbar^2 e^2\over 2m} \right) 
\ee  
and
$$ 
S_I(E2,A_1+A_2)= $$
\be
\left( {25\over 2\pi} \right){1\over Z}  
\left(Z_1Z_2 +\Bigl(Z_1{A_2\over A}-Z_2{A_1\over A}\Bigr)^2 \right) S_0^2 
\left( {\hbar^2 e^2\over 2m} \right) \,,
\ee 
under the assumption  that the nucleus with mass $A$ and charge $Z$ is  
split in two clusters with masses $A_1$ and $A_2$, charges $Z_1$ and $Z_2$ and 
neutron numbers $N_1$ and $N_2$. The distance 
$S_0$ is the equilibrium separation that may be simply calculated as the sum 
of the radii of the two clusters (we have taken $S_0=3.63$ fm). We find 
that the low-lying dipole strength exhausts approximatively the 2.6\% of 
the Thomas-Reiche-Kuhn sum rule, but it amounts to about 94\% of the energy  
weighted molecular dipole sum rule. Similarly the quadrupole strength is the 
9.2\% of the energy weighted quadrupole sum rule and about 42\% of the 
EWMSR. For a proper comparison with the sum rule, 
we have included 
in the calculation of the exhausted fraction 
of sum rules, besides the transition to the continuum and 
the quadrupole transition to the  
first excited state, 
all the possible transitions to lower unphysical bound states 
($1s_{1/2},2s_{1/2},1d_{5/2},1d_{3/2}$ for dipole and $1p_{3/2},1p_{1/2}$ for 
quadrupole). Note that the dipole transitions to unphysical 
states give a negative contribution of about  
50 \% to the total energy weighted strength.

\section{Formalism and Form Factors} 
We move now from the pure electromagnetic response to the study of  
the breakup reaction in which the dicluster $^{7}$Li 
nucleus is used as a projectile on a heavy target. 
 
The coordinate system for the interaction between a dicluster 
nucleus and a target is depicted in fig. \ref{coord}. 
The factors $f_1$ and $f_2$ are the ratios of the distances of the center of  
mass of each cluster from the common center of mass divided by the  
inter-cluster 
distance $r$. We have named the two clusters as 'core' and 'cluster' to avoid 
confusions even if the alpha particle has not a mass large enough  
to justify the  
choice with respect to the triton. 

\begin{figure*}[!t]
\epsfig{figure=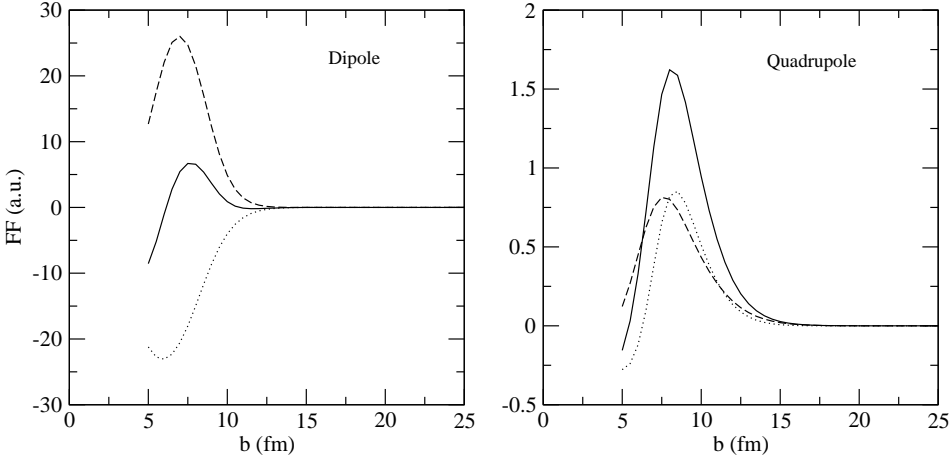,width=.7\textwidth} 
\caption{Form factors for dipole and quadrupole cases as a function of the
distance $r$ for the transition $p_{3/2,3/2}\rightarrow s_{1/2,1/2}$ for
an excitation energy in the continuum $E_C=1$ MeV. The three different 
curves correspond to $~^7$Li-Target (solid line), $\alpha$-Target (dotted)
and $t$-Target interactions.} 
\label{canc} 
\end{figure*} 

Within the cluster model, the wave function for $^{7}$Li is the product 
of the  
wave functions of the two alpha and triton clusters (assumed to be frozen  
during the transition) and of the wave function 
describing the relative cluster-cluster  
motion. Assuming a value L for the relative angular momentum, and taking  
into account the 
intrinsic j=0 and j=1/2 spins for the alpha and the triton, the generic state 
with total spin J can be expressed as $\mid L,J,M \rangle$.  For states in  
the continuum, 
states are also characterized by the value $E_C$ of the energy 
in the continuum, namely $\mid L,J,M ; E_C \rangle$.  The formfactor 
associated with a process in which the relative cluster-cluster motion  
undergoes a transition to the continuum is given by 
$$ 
F(\vec R,E_C)_{LJM\rightarrow L'J'M';E_C}  \=$$
\be
\= \langle L'J'M' ;E_C \mid V (\vec R, \vec r) \mid LJM \rangle \ 
\ee 
in terms of the relative projectile-target coordinate $\vec R$. 
The relevant interaction is assumed as the sum of the interactions of  
the target, labeled with $T$, with each cluster,  
\be 
V (\vec R, \vec r)=  V_{\alpha-T}
(\mid \vec R-f_2 \vec r\mid)+V_{t-T}(\mid\vec R+f_1 \vec r\mid)\,, 
\ee 
where each interaction consists in a nuclear and a Coulomb part, the former 
being assumed to be of a Woods-Saxon form.
Since the clusters are frozen during the transition   
the integration over the internal degrees 
of freedom is straightforward and one is left with an integration over the  
cluster-cluster coordinate $\vec r$ in the form 

\begin{gather}
F(\vec R,E_C)_{LJM\rightarrow L'J'M';E_C}\=  \sqrt{\pi}\hat J\hat J' \hat L \hat L'\nonumber\\
 \sum_{\lambda,\mu} (-1)^{3j_{cl}-M'}
\Biggl( {J \atop -M} {J' \atop M'} {\lambda \atop \mu} \Biggr)  
\Biggl( {L' \atop 0} {L \atop 0} {\lambda \atop 0} \Biggr)  
\Biggl\{ {J' \atop L}{J \atop L'}{\lambda \atop j_{cl}} \Biggr\} \nonumber \\
\Biggl[ \int_0^\infty r^2 dr \int_{-1}^1 du \psi_{L}(r) \psi _{L',E_C}(r) 
\Bigl( V_{\alpha-T}(\mid \vec R-f_2 \vec r\mid)+  \nonumber \\
V_{t-T}(\mid\vec R+f_1 \vec r\mid) \Bigr) 
 P_{\lambda}(u)\Biggr] Y_{\lambda,\mu} (\hat R)  \,,
\label{fofa} 
\end{gather} 
where ${\lambda,\mu}$ are the change in orbital angular momentum and its  
third component due to the transition and $u$ is the cosine of the angle  
between the two vectors $\vec R$ and $\vec r$. As in the previous section, 
$\hat j=(2j+1)^{1/2}$.
 
The resulting Coulomb and nuclear form factors for the $^{7}$Li +  
$^{165}$Ho reaction are plotted 
in fig. (\ref{formfact})   
for a dipole transition between  the $p_{3/2}$   
ground state and the  
$s_{1/2}$ state at $E_C =1$ MeV in the continuum. It is evident that  
the nuclear field 
dominates at smaller distances, while the Coulomb one dominates at larger  
distances. This is once again displayed in the next three figures where three 
different distances have been kept constant and the Q-value dependence upon  
$E_C$ is illustrated. 
The nuclear contribution is still very important at a distance of $12-14$ fm 
that is far beyond the geometrical sum of the radii of the two systems.  
This effect may even be magnified   
in halo systems closer to the drip lines, where the weakly-bound  
wavefunctions are even more extended.   
Figures of qualitatively similar behaviour are obtained for all the other
possible transitions.  
 
To better understand the relative role of dipole and quadrupole interactions,  
we show separately in fig. \ref{canc} 
the form factors for selected dipole and quadrupole transitions. 
Together with the total form factor, we report the contribution  
arising from the interaction between each of the two clusters and the target  
separately. In the case of a dipole transition 
a cancellation occurs between the two contributions, while for the quadrupole 
case the two clusters contribute constructively to the excitation.  The   
effect is here amplified by the fact that the two clusters have similar sizes. 
In the limit of two equal clusters, the nuclear contribution to dipole  
transitions 
would exactly vanish.  Similarly, no Coulomb dipole transitions are allowed 
if the two clusters have equal mass to charge ratios.

\section{Cross section} 
The formfactors obtained in the last section contain all the  
relevant elements to  
build up breakup cross-sections and Q-value distributions in a simple, 
although accurate procedure. 
The reaction amplitudes can be calculated in a semiclassical coupled-channel  
approach. 
\begin{figure*}[!t]
\epsfig{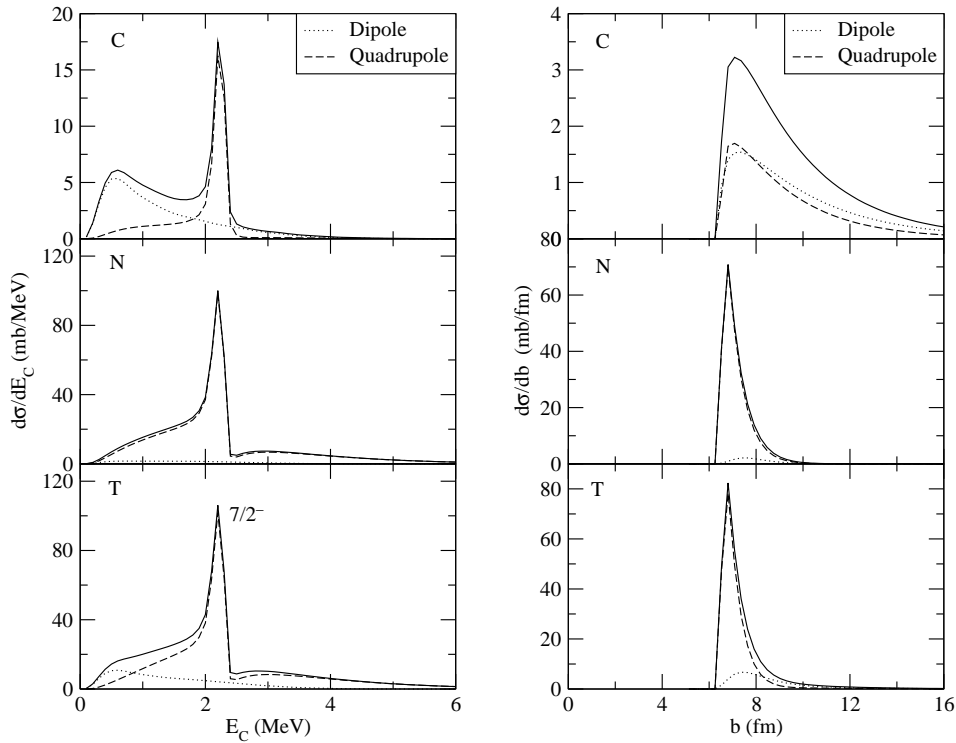} 
\caption{Left panels: Q-value distribution for (Coulomb, Nuclear and 
Total or Interference) breakup of $^7$Li on $^{165}$Ho at $E_{cm}=40$ MeV. 
Dipole (dotted) and quadrupole (dashed) contributions are shown together 
with their sum (full line). The $7/2^-$ resonance is marked, 
while the $5/2^-$ around 4 MeV is very small. 
Right panels: Differential (Coulomb, Nuclear and Total) breakup cross 
sections as a function of the impact parameter with the same data of the 
left figures. Again different multipolarities are shown separately and one 
may notice the different behaviour at large impact parameter, that is 
dominated by the dipole contribution.} 
\label{cross} 
\end{figure*}      
The energy in the continuum is divided in a suitable number of intervals,  
treated as different channels. For each energy interval (and each spin),  
the formfactor connecting with the ground state is obtained 
(assuming the central value of the energy interval) 
as described in the previous section.  To keep simple the calculations
continuum-continuum couplings have not been included, although in some cases 
they were found to play a relevant role (see for example ref. \cite{Filo}). 
We follow in time the solutions
of the system of coupled equations for the amplitudes in the different
channels, along a  trajectory that is calculated semiclassically using 
a standard Akyuz-Winther parameterization \cite{Brog} for the 
target-projectile potential.
The values of the amplitudes at the end of the scattering process 
are then used to calculate cross sections for the excitation  
of a given channel and differential cross sections as a function of the  
energy in the continuum. The simplicity of the scheme allows us to use
a rather small energy mesh, without any convergence problem as the
mesh is reduced. This is particularly important when one has to deal with 
a continuum that includes, as in our case, narrow resonances. Details
on the formalism may be found in Ref. \cite{Das2}.
The resulting cross sections are collected in Fig. \ref{cross}.  
The Q-value distribution obtained for Coulomb  
breakup is displayed in left-upper panel (the contributions of the dipole  
and quadrupole transitions are separately shown, together with their sum). 
It is worthwhile noticing that the two low-lying peaks arise from  
different mechanisms: the peak at around $0.5$ MeV is mostly build up 
with transition 
to the continuum which are enhanced due to the weak-binding nature 
of the projectile, while the peak at $2.186$ MeV has a true 
resonant nature ($7/2^-$).  In the case of quadrupole transition
the non-resonant strength can be less easily seen just above the threshold, 
since the different radial dependence of the formfactor makes 
its relative magnitude   
small compared to the dipole one, in spite of a larger value of the 
$B(E\lambda)$'s distribution. Differential cross sections with respect 
to impact parameter are also shown in the right column of fig. \ref{cross}.
 At low impact  
parameters the excitation process is strongly quenched by the 
transmission factor. For large values 
one can see the different behaviour of the two tails: 
the quadrupole contribution decays faster than the 
dipole. Consequently at large impact parameters, that, 
in a classical picture, correspond to forward angles, 
the Coulomb breakup cross sections are mostly due to dipole 
transitions to the continuum.   
The total Coulomb cross section (resonant and non-resonant) 
at $E_{cm}=40$ MeV amount to $\sim 4.85$ mb, with comparable 
dipole and quadrupole 
contributions ($\sim 3.0$ mb for the dipole and $\sim 1.85$ mb for the 
quadrupole).  

To evaluate the effect of the nuclear interaction we need to specify 
the precise set of optical parameters between each of the two clusters
and the target nucleus. We have looked up for optical parameters in the 
standard tables \cite{alfa,trit}, where
data sets for elastic scattering on holmium are missing, and, in absence
of any alternative, we have used parameters extracted for cerium, 
which is the closest isotope. We used only the real part of the potentials
in the construction of the form factors.
Of course we do not expect these 
parameters to represent a strictly valid quantitative choice, but we have 
used them in order to give estimates of the nuclear and total breakup cross 
sections that are reported in the following.
The nuclear breakup has a Q-value distribution 
(depicted in fig. \ref{cross}, second row of the 
left column) with an overall profile that resembles the Coulomb one,
being the total integrated cross-section about $\sim 24.1$ mb.  
At a variance with previous findings the dipole  
contribution to this cross-section ($\sim 1.3$ mb) is now much smaller 
that the quadrupole  
one ($\sim 22.8$) mb.  This is again originated by the comparable size 
of the clusters, that 
hinders dipole components, while the predominant nuclear quadrupole 
term is not quenched by the faster radial dependence of the formfactor 
as in the case of the Coulomb term.   
In this case, therefore, both resonant and non-resonant peaks are of
predominant quadrupole nature. 
 
The final Q-value distribution and the corresponding curve as a function 
of the impact parameter, which both take into account the interference 
between the two fields, are depicted in the last row of fig. 
\ref{cross}. The total cross sections amounts to about $\sim 29.0$ mb.  
The dipole transition contributes for $\sim 6.9$ mb, while the quadrupole  
is about $\sim 22.1$ mb.   
It should be noticed that, while the Coulomb contribution is rather 
insensitive to the absorption radius, a significant change in the nuclear  
cross-section (and therefore in the total) may occur as long as the  
radius of the nuclear interaction is 
varied, as as one can easily infer, for example, from the last panel. 
For a discussion on the subject, see for example  ref. \cite{Das2}. 
 
In order to test our model and to compare with other available models
we have performed calculations of breakup cross sections for $~^7$Li on
$~^{208}$Pb at 48 MeV bombarding energy. Our calculations may be directly
confronted with the work of Kelly and collaborators \cite{Kelly}, that are 
essentially based on the same physical ingredients.
We report both theirs and our results in Table II. 
Standing the differences in the values of the couplings 
(we extracted $~^{208}$Pb-$\alpha$ and $~^{208}$Pb-$t$ optical parameters 
from the work of Gupta {\it et al.} \cite{Gupta}) and in the treatment of the 
continuum (we take into account continuum energy up to 10 MeV),
the agreement among the various contributions to the cross sections from 
$\lambda=0,1$ and $3$ states is satisfactory, although some discrepancy is
seen in the $\lambda=3$ continuum. In addition we provide 
calculations for the contribution arising from $d$ states, that is found to
be a very important component of the total cross section. This is at 
variance with respect to the cited analysis by Kelly {\it et al.},
where the quadrupole component is considered to be negligible. 
The numerical results
in Table II have been obtained considering dipole and quadrupole transitions 
only, but we have checked that octupole transitions to $d$ states and 
hexadecupole transitions to $f$ states may be neglected (being around 
10$^{-3}$mb and 0.9 mb respectively).

\begin{table}[t] 
\begin{center} 
\begin{tabular}{|c|c|c|} 
\hline 
 & \multicolumn{2}{|c|}{$\sigma(~^7Li\rightarrow \alpha+t)$ (mb)}\\
$L'(\hbar)$& ~~Ref.\cite{Kelly}~~ & This work \\ \hline
 0 & 26.3 & 22.9 \\
 1 &  6.0 &  4.4 \\
 2 &   -  & 25.3 \\
 3 & 15.9 & 6.7  \\ \hline
Total& 48.2 & 59.3\\ \hline
\end{tabular}
\end{center}
\caption{Cross sections for the breakup of $~^7$Li into $\alpha + t$ cluster 
states in the $~^7$Li $+~^{208}$Pb reaction at $E=48MeV$. 
The contributions arising from final states with given 
angular momentum ($L'$) are separately listed.}
\end{table}

\section{Conclusions} 
We have illustrated a general model to describe excitations to continuum 
states in 
weakly-bound dicluster nuclei, leading to cluster dissociation.  In the model  
the internal degrees of freedom of the clusters are kept frozen in the 
excitations, which are therefore entirely ascribed to 
the relative cluster-cluster motion.  Both resonant and non-resonant 
continuum states are simultaneously properly included. 
In the case of weakly-bound nuclei 
the non-resonant part shows the presence of multipole strength  
at the threshold that is a typical feature for  
single-particle excitations in one-particle halo nuclei. 

Paralleling the formalism previously developed for the break-up of 
one-particle halo nuclei,  
formfactors for transitions to cluster continuum states are constructed and 
cross sections for cluster break-up reactions are calculated in a 
semiclassical coupled 
channel description.  The interplay of dissociation via resonant states or via 
non-resonant continuum is discussed. 
The formfactors are studied in detail: we illustrate their behaviour as a 
function of the relative distance of the two colliding
nuclei, as a function of the energy of the continuum states and we 
discuss the effects of cancellation and reinforcement, for dipole and 
quadrupole transitions respectively, that is 
a consequence of the relative masses and charges of the two clusters.

Q-value distributions and differential break-up cross-sections with respect 
to impact parameters, as well as the total break-up cross-sections, are
evaluated for the reaction $^7$Li$+^{165}$Ho, taking into account both nuclear
and Coulomb contributions,
although restricted to dicluster nuclei. This simple approach to the 
break-up problem avoids possible problems arising from 
a crude energy binning of the continuum.
Our results show that, in the case of a system described in terms of two 
clusters of similar size and charge (as it is the case of $~^7$Li), 
the main contribution to break-up 
processes come from the nuclear quadrupole mechanism.
 
\begin{acknowledgement} 
The authors acknowledge fruitful discussions with K.Hagino and C.H.Dasso.
\end{acknowledgement}

\end{document}